\documentclass[pra,10pt,twocolumn,superscriptaddress,showpacs]{revtex4-1}
\usepackage{amsmath}
\usepackage{latexsym}
\usepackage{amssymb}
\usepackage[colorlinks=true, citecolor=blue, urlcolor=blue]{hyperref}
\usepackage{float}
\usepackage{graphics}
\usepackage[pdftex]{graphicx}
\newcommand{\ket}[1]{| #1 \rangle}
\newcommand{\bra}[1]{\langle #1 |}

\begin{document}

\title{Distinguishing different classes of entanglement of three-qubit pure states}
\author{Chandan Datta}
\email{chandan@iopb.res.in}
\affiliation{Institute of Physics, Sachivalaya Marg, Bhubaneswar 751005, Odisha, India.}
\affiliation{Homi Bhabha National Institute, Training School Complex, Anushakti Nagar, Mumbai 400085, India.}
\author{Satyabrata Adhikari}
\email{satyabrata@dtu.ac.in} \affiliation{Delhi Technological University,
Shahbad Daulatpur, Main Bawana Road,
Delhi 110042, India.}
\author{Arpan Das}
\email{arpandas@iopb.res.in}
\affiliation{Institute of Physics, Sachivalaya Marg, Bhubaneswar 751005, Odisha, India.}
\affiliation{Homi Bhabha National Institute, Training School Complex, Anushakti Nagar, Mumbai 400085, India.}
\author{Pankaj Agrawal}
\email{agrawal@iopb.res.in} \affiliation{Institute of Physics,
Sachivalaya Marg, Bhubaneswar 751005, Odisha, India.}
\affiliation{Homi Bhabha National Institute, Training School Complex, Anushakti Nagar, Mumbai 400085, India.}

\begin{abstract}
Employing the Pauli matrices, we have constructed a set of operators,
which can be used to distinguish six inequivalent classes of
entanglement under SLOCC (stochastic local operation and classical communication) 
for three-qubit pure states. These operators have very simple structure and can
be obtained from
the  Mermin's operator with suitable choice of directions.
Moreover these operators may be implemented in an
experiment to distinguish the types of entanglement present in a
state. We show that the measurement of only one operator is
sufficient to distinguish GHZ class from rest of the classes.
It is also shown that it is possible to detect and classify other classes by performing 
a small  number of measurements. We also show how to construct such observables
in any basis. We also consider a few mixed states to investigate the usefulness of our operators.
Furthermore,  we consider the teleportation scheme of Lee et al. \cite{lee}
and show that the partial tangles and hence teleportation fidelity can be measured. We have also shown that these partial
tangles can also be used to classify
genuinely entangled state, biseparable state and separable state.
\end{abstract}


\maketitle


\section{Introduction}

Entanglement is one of the key features of quantum mechanics, which differentiates quantum world from the classical world. It is an essential resource
for many information processing tasks, such as quantum cryptography \cite{1}, teleportation \cite{2}, super-dense coding \cite{3,4} etc. Also from the
foundational perspective of quantum mechanics, entanglement is unparalleled for its supreme importance. Therefore, its characterization and
quantification is very important from both theoretical as well as experimental point of view.

A lot of research have been carried out to
understand entanglement qualitatively and quantitatively.
Quantification and characterization of entanglement is unambiguous for pure bipartite state, but not for the mixed states \cite{5,10,chandan}.
In the spirit of resource theory of entanglement \cite{5,6}, two entangled states are said to be equivalent if they can be obtained from each
other with certainty with respect to LOCC  (local operation and classical communication). Entanglement of any pure bipartite state is uniquely captured by the entropy of entanglement in the asymptotic limit \cite{7}. But this is not true for mixed states. There is no unique quantification of
entanglement for this case and a number of entanglement measures and monotones \cite{10} have been constructed over the years. Situation gets worse
for multipartite scenario, both for pure and mixed states. One can straightforwardly extend some of the entanglement measures and monotones constructed
for bipartite systems to multipartite scenario, but there is no unique quantification of entanglement in multipartite scenario even for pure states.
We can not even define a unique maximally entangled multipartite state and there are many inequivalent forms of entanglement in the asymptotic
limit \cite{8,9}. In this paper,  we will restrict ourselves to the entanglement properties of a single copy of a pure multipartite state. For a single copy, two states are LOCC
equivalent if and only if they are related by LU (local unitary) \cite{9,11}. But in the single copy restriction, even two bipartite pure states
are not typically related by LU. To evade this difficulty, the LOCC operation, through which the conversion of entangled states is considered is
slightly loosened. One now considers the conversion of states through stochastic local operation and classical communication (SLOCC), i.e two entangled
states are converted to each other by means of LOCC but with a non-vanishing probability of success \cite{9}. Two states are now called SLOCC
equivalent if they can be obtained from each other under SLOCC, otherwise they are SLOCC inequivalent. For three-qubit pure states all possible SLOCC
inequivalent classes have been characterized \cite{12}. There are six SLOCC inequivalent classes: separable, 
three bi-separable and two genuinely entangled (GHZ and W).
In general it is very difficult to characterize and distinguish different classes from each other. 
For three-qubit pure states analytical
characterization is present in literature using local entropies and the concept of tangle \cite{12}. But from an experimental point of view these are not realizable.

In this paper, we will be providing an feasible way
to distinguish these six different classes of pure three-qubit
states. In a very recent paper \cite{13}, a proposed set of Bell
inequalities can distinguish separable, biseparable and genuine
entanglement for pure three-qubit states by the pattern of
violations of the Bell inequalities within the set. In another
work,  Zhao et.al. \cite{zhao} have provided the necessary and
sufficient conditions to classify the separable, biseparable and
genuine entangled state. But they did not succeed in distinguishing the
GHZ-type and W-type states, which fall under the category of
genuine entangled state. They have shown that
measurement of seven observables are needed to
 classify and detect pure three-qubit entangled state.

In this work, we will not consider
Bell inequalities but will construct operators, which can distinguish
six classes of entanglement for pure three-qubit states.
Firstly, we have related the tangle of the state with the
expectation value of one operator, which consist of Pauli
operators. This means that we may experimentally determine the
value of the tangle. Non zero value of the tangle guarantees that
the given three-qubit pure state is a GHZ type state. Therefore,
in our case only one operator is needed to differentiate between
GHZ-type state and the other classes of three-qubit pure states.
Measurement of three operators are needed to distinguish W-type state
and biseparable/separable states. Measurement of four more operators
are needed to classify biseparable and separable states. But Zhao
et.al. \cite{zhao} have shown that measurement of seven operator
is needed to classify the pure three-qubit state.

One can transform a given pure three-qubit  state  to the
canonical form  \cite{acin,acin1}. We have constructed the operators and proved our
results using this form of an arbitrary state. In some experiments, the states
are created in this canonical form \cite{nmr}, then our list of observables 
can be used without a change.  But  most of the time this is not the case,
and a state is generated in a different basis.
In such a case, one can find  another set of suitable observables by applying
local unitary transformations on the given set of observables.
 We explicitly demonstrate this by considering
two states in computational basis. This is possible because different sets of
basis vectors are connected by unitary transformations. If we know the two sets,
one can construct suitable unitary transformations.  However, there is a drawback.
Sometime, one has
to know the state for constructing suitable unitary transformations, as they
 depend on the state parameters. Although our scheme is about distinguishing
 various SLOCC classes of pure states, we also examine the usefulness of our 
 operators in the context of a few mixed states.

We also consider the teleportation scheme of Lee et al. \cite{lee}. They introduced the concept of partial tangle and related the fidelity
of their teleportation scheme to these partial tangles. We relate
these partial tangles to a set of observable quantities. By measuring
these quantities, one can find out the values of partial tangles and
the fidelity of the teleported state. In this way,  the partial
tangles can be also be used to classify
genuinely entangled state, biseparable state and separable state

We have organized the paper as follows. In the section 2,
 we have introduced the idea of tangle and its observable measure. In the section 3,
the classification of different classes are done. Implementation of our scheme in
 any other basis has been discussed in the section 4. We have discussed the usefulness of our measurement observable for a few class of mixed states in the section 5. In the section 6, an experimental way of measuring fidelity for a teleportation scheme has been
discussed using same kind of operators. Finally, we conclude in the last section.

\section{Tangle and it's observable measure}

Tangle was first introduced in \cite{tangle} in the context of
distributed entanglement, to quantify the amount of three-way
entanglement in a three-qubit state. For a pure state it can be
interpreted as residual entanglement, which is not captured by
two-way entanglement between the qubits. It was also shown to be
an entanglement monotone \cite{12}. The tangle is defined as
\begin{equation}\label{tangle}
\tau=C_{A(BC)}^2-C_{AB}^2-C_{AC}^2,
\end{equation}
where $C_{AB}$ and $C_{AC}$ denote the concurrence
\cite{concurrence} of the entangled state between the qubits $A$
and $B$ and between the qubits $A$ and $C$ respectively.
The concurrence $C_{A(BC)}$ refers to the entanglement of qubit $A$ with the joint state of qubits $B$ and $C$.

Any three-qubit pure state can be written in the canonical form
\cite{acin,acin1},
\begin{equation}\label{three-qubit state}
\ket{\psi}=\lambda_0\ket{000}+\lambda_1 e^{i
\theta}\ket{100}+\lambda_2\ket{101}+\lambda_3\ket{110}
+\lambda_4\ket{111},
\end{equation}
where $\lambda_i\geqslant 0$, $\sum_i \lambda_i^2=1$, $\theta \in
[0,\pi]$ and $\{{\ket{0},\ket{1}}\}$ denote the
basis of Alice's, Bob's and Charlie's Hilbert space. The tangle
for the state $\ket{\psi}$ given in (\ref{three-qubit state}) is
found to be
\begin{equation}\label{tangle three-qubit}
\tau_{\psi}=4\lambda_0^2\lambda_4^2.
\end{equation}
The tangle as given in (\ref{tangle three-qubit}) may be measured experimentally if we take the expectation value of the operator
\begin{equation}\label{tangle operator}
O= 2 (\sigma_x\otimes\sigma_x\otimes\sigma_x),
\end{equation}
with respect to the state $\ket{\psi}$. The operator $O$ given in
(\ref{tangle operator}) can be obtained by suitably choosing the
unit vectors in Mermin operator, which is defined as
\cite{chi,satya}
\begin{eqnarray}\label{mermin operator}
B_M&=&\hat{a}_1.\vec{\sigma}\otimes\hat{a}_2.\vec{\sigma}
\otimes\hat{a}_3.\vec{\sigma}-\hat{a}_1.\vec{\sigma}\otimes\hat{b}_2.\vec{\sigma}
\otimes\hat{b}_3.\vec{\sigma}-\nonumber\\
&&\hat{b}_1.\vec{\sigma}\otimes\hat{a}_2.\vec{\sigma}
\otimes\hat{b}_3.\vec{\sigma}-\hat{b}_1.\vec{\sigma}\otimes\hat{b}_2.\vec{\sigma}
\otimes\hat{a}_3.\vec{\sigma},
\end{eqnarray}where $\hat{a}_j$, $b_j$ $(j=1,2,3)$ are the measurement direction for the $j$th party and $\vec{\sigma}=(\sigma_x,\sigma_y,\sigma_z)$
are the usual Pauli matrices. By choosing the unit vectors as
$\hat{a}_1=(1,0,0)$, $\hat{a}_2=(1,0,0)$, $\hat{a}_3=(1,0,0)$,
$\hat{b}_1=(-1,0,0)$, $\hat{b}_2=(1,0,0)$ and $\hat{b}_3=(1,0,0)$,
we can construct the operator $O$. Therefore, the expectation
value of the operator $O$ in the state $\psi$ is given by
\begin{equation}\label{tangle observable measure}
\langle O \rangle_{\psi}=\langle \psi |O|\psi \rangle=
4 \lambda_0\lambda_4=2\sqrt{\tau_{\psi}}.
\end{equation}
Hence, from (\ref{tangle observable measure}) it is clear that by measuring the expectation value of $O$, one can easily calculate the value of the tangle.


\section{Classification of Three-Qubit Pure States}
In this section, we will show how to classify six different
classes of three-qubit pure states. It is known that
tangle is nonzero only for GHZ class \cite{12}; it is zero for other five
classes. So using (\ref{tangle observable measure}) one
can separate GHZ class from other five classes.
Since it is not possible to distinguish zero tangle classes of
three-qubit pure states with a single quantity $\tau_{\psi}$,  so we
need to define other observables. To fulfill our aim, let us
consider two quantities $P$ and $Q$, which can be defined as
\begin{equation}\label{P}
P=\langle\psi|O_1|\psi\rangle \langle\psi|O_2|\psi\rangle=\langle O_1 \rangle_{\psi} \langle O_2 \rangle_{\psi},
\end{equation}
and
\begin{equation}\label{Q}
Q=\langle O_1 \rangle_{\psi}+\langle O_2 \rangle_{\psi}
+\langle O_3 \rangle_{\psi}.
\end{equation}
The operators $O_1$, $O_2$ and $O_3$ are given by
\begin{equation}\label{O1}
O_1=2 (\sigma_x\otimes\sigma_x\otimes\sigma_z),
\end{equation}
\begin{equation}\label{O2}
O_2=2 (\sigma_x\otimes\sigma_z\otimes\sigma_x)
\end{equation}
and
\begin{equation}\label{O3}
O_3=2 (\sigma_z\otimes\sigma_x\otimes\sigma_x).
\end{equation}
The operator $O_1$ given in (\ref{O1}) can be obtained from the Mermin operator (\ref{mermin operator}) by choosing the unit vectors as $\hat{a}_1=(1,0,0)$, $\hat{a}_2=(1,0,0)$, $\hat{a}_3=(0,0,1)$, $\hat{b}_1=(-1,0,0)$, $\hat{b}_2=(1,0,0)$ and $\hat{b}_3=(0,0,1)$. One can find operator $O_2$ given in (\ref{O2}) by choosing the unit vectors as $\hat{a}_1=(1,0,0)$, $\hat{a}_2=(0,0,1)$, $\hat{a}_3=(1,0,0)$, $\hat{b}_1=(-1,0,0)$, $\hat{b}_2=(0,0,1)$ and $\hat{b}_3=(1,0,0)$. Similarly operator $O_3$ given in (\ref{O3}) can be obtained by choosing the unit vectors as $\hat{a}_1=(0,0,1)$, $\hat{a}_2=(1,0,0)$, $\hat{a}_3=(1,0,0)$, $\hat{b}_1=(0,0,-1)$, $\hat{b}_2=(1,0,0)$ and $\hat{b}_3=(1,0,0)$. The expectation value of the operators $O_1$, $O_2$ and $O_3$ with respect to the state $\ket{\psi}$ are as follows,
\begin{eqnarray}\label{exp O1O2O3}
&&\langle O_1 \rangle_{\psi}=4 \lambda_0\lambda_3,\nonumber\\&&\langle O_2 \rangle_{\psi}=4 \lambda_0\lambda_2 \quad \mbox{and}\nonumber\\
&&\langle O_3 \rangle_{\psi}=-4 (\lambda_2\lambda_3+\lambda_1\lambda_4\cos\theta).
\end{eqnarray}
Therefore, using (\ref{exp O1O2O3}), we can obtain $P$ and $Q$ as
\begin{eqnarray}\label{exp PQ}
&&P=16 \lambda_0^2\lambda_2\lambda_3 \quad \mbox{and}\nonumber\\&&Q=4\Big(\lambda_0\lambda_3+
\lambda_0\lambda_2-
(\lambda_2\lambda_3+\lambda_1\lambda_4\cos\theta)\Big).
\end{eqnarray}
We are now in a position to classify zero tangle three-qubit pure
states based on the expectation values of the operators $O_1$,$O_2
$,$O_3$ and the two quantities $P$ and $Q$.\\

\textbf{Theorem 1:} Any three-qubit state belongs to the $W$ class if,
\begin{eqnarray}\label{W class prop}
&& (i) \tau_{\psi}=0,\nonumber\\&&
(ii) P\neq 0.
\end{eqnarray}

\textbf{Proof :} Using parametrization (\ref{three-qubit state}), any three-qubit pure state, which is in $W$ class can be written as \cite{acin1,12},
\begin{equation}\label{W state}
\ket{\psi}_W=\lambda_0\ket{000}+\lambda_1 \ket{100}+\lambda_2\ket{101}+\lambda_3\ket{110}.
\end{equation}
As there is no $\lambda_4$, so from (\ref{tangle three-qubit}) it is clear that $\tau_{\psi_W}=0$. From (\ref{exp PQ}) one can find $P=16 \lambda_0^2\lambda_2\lambda_3 \neq 0$ and $Q=4(\lambda_0\lambda_3+
\lambda_0\lambda_2-
\lambda_2\lambda_3) \neq 0$.

We will deduce the conditions by which it is possible to distinguish three biseparable classes.\\

\textit{Lemma 1 :} Any three-qubit state is biseparable in $1$ and $23$ bipartition if
\begin{eqnarray}\label{bi 1/23 class prop}
&& (i) \tau_{\psi}=0,\nonumber\\&&
(ii) \langle  O_1 \rangle_{\psi}=0,\nonumber\\&&
(iii) \langle O_2 \rangle_{\psi}=0 \quad \mbox{and}\nonumber\\&&
(iv) \langle O_3 \rangle_{\psi} \neq 0.
\end{eqnarray}

\textit{Proof :} Any pure three-qubit state which is biseparable in 1 and 23 bipartition, can be written as $\ket{0}(\alpha\ket{00}+\beta\ket{11})$, upto some local unitary transformation \cite{12}. Canonical form of three-qubit pure states as written in (\ref{three-qubit state}) will have the aforesaid biseparable form if all the $\lambda_i$'s except $\lambda_1$ and $\lambda_4$ are zero. Hence, the state belonging to $1$ and $23$ bipartition can be written in terms of $\lambda_i$'s as
\begin{equation}\label{bi 1/23 state}
\ket{\psi}_{1|23}=\ket{1}(\lambda_1 \ket{00}+
\lambda_4\ket{11}).
\end{equation}

As $\lambda_0=0$, the tangle is zero for this class of state. From (\ref{exp O1O2O3}) we notice that, $\langle O_1 \rangle_{\psi}=0$, $\langle O_2 \rangle_{\psi}=0$ and $\langle O_3 \rangle_{\psi}=
-4\lambda_1\lambda_4$. Hence, $P = 0$ and $Q \neq 0$.\\

\textit{Lemma 2 :} Any three-qubit state is biseparable in $12$ and $3$ bipartition if,
\begin{eqnarray}\label{bi 12/3 class prop}
&& (i) \tau_{\psi}=0,\nonumber\\&&
(ii) \langle O_1 \rangle_{\psi}\neq 0,\nonumber\\&&
(iii) \langle O_2 \rangle_{\psi}=0 \quad \mbox{and}\nonumber\\&&
(iv) \langle O_3 \rangle_{\psi} = 0.
\end{eqnarray}

\textit{Proof :} The state, which belongs to $12$ and $3$ bipartition can be written as
\begin{equation}\label{bi 12/3 state}
\ket{\psi}_{12|3}=(\lambda_0\ket{00}+\lambda_3\ket{11})\ket{0}.
\end{equation}
The tangle is zero as $\lambda_4=0$. Using (\ref{exp O1O2O3}) we can infer that $\langle O_1 \rangle_{\psi}=4\lambda_0\lambda_3$, $\langle O_2 \rangle_{\psi}=0$ and $\langle O_3 \rangle_{\psi}=0$. Therefore, $P = 0$ and $Q \neq 0$. \\

\textit{Lemma 3 :} Any three-qubit state is biseparable in $13$ and $2$ bipartition if
\begin{eqnarray}\label{bi 13/2 class prop}
&& (i) \tau_{\psi}=0,\nonumber\\&&
(ii) \langle O_1 \rangle_{\psi}= 0,\nonumber\\&&
(iii) \langle O_2 \rangle_{\psi} \neq 0 \quad \mbox{and}\nonumber\\&&
(iv) \langle O_3 \rangle_{\psi} = 0.
\end{eqnarray}

\textit{Proof :} The state belongs to $13$ and $2$ bipartition can be written as
\begin{equation}\label{bi 13/2 state}
\ket{\psi}_{13|2}=\lambda_0\ket{000}+\lambda_2\ket{101}.
\end{equation}
The tangle is zero as $\lambda_4=0$. The expectation values of the operators $O_1$, $O_2$ and $O_3$ in this state are as follows $\langle O_1 \rangle_{\psi}=0$, $\langle O_2 \rangle_{\psi}=4\lambda_0\lambda_2$ and $\langle O_3 \rangle_{\psi}=0$. Therefore, $P = 0$ and $Q \neq 0$.

We can now use these lemmas to prove the following theorem.\\

\textbf{Theorem 2:} Any three-qubit pure state is biseparable if,
\begin{eqnarray}\label{bi class prop}
&& (i) \tau_{\psi}=0,\nonumber\\&& (ii) P=0 \quad
\mbox{and}\nonumber\\&& (iii) Q \neq 0.
\end{eqnarray}

\textbf{Proof :} From the above lemmas, it is clear that for a biseparable state,
either $ \langle O_1 \rangle_{\psi}= 0$, or $ \langle O_2 \rangle_{\psi}= 0$. As
$P$ is the product of these two expectation values, therefore $P =0$ for any 
biseparable three-qubit pure state. The quantity $Q$ is the sum of the expectation
values of the operators $O_1, O_2$ and $O_3$, and according to the above three
lemmas, at least one is nonzero. Therefore $Q \neq 0$. This proves the theorem.\\

\textbf{Theorem 3:} Any three-qubit state is separable if
\begin{eqnarray}\label{sep class prop}
&& (i) \tau_{\psi}=0,\nonumber\\&&
(ii) P=0 \quad \mbox{and}\nonumber\\&&
(iii) Q = 0.
\end{eqnarray}

\textbf{Proof :} Any separable three-qubit pure state can be written as $\ket{0}\ket{0}\ket{0}$, after applying some appropriate local unitary operation. For this state $\tau_{\psi}$, $P$ and $Q$ all are zero. That completes our proof.\\

From the above theorems and lemmas we can classify all the classes of three-qubit pure states. Moreover, as these observables only contain Pauli matrices, they can be measured in experiments. We note that there is some arbitrariness in the definition of $P$. Above proofs will go through, even if we
would have have defined $P$ as a product of the expectation values of operators ``$O_2$ and $O_3$'',
or ``$O_1$ and $O_3$'', instead of operators ``$O_1$ and $O_2$''.


\section{Local unitary equivalence with computational basis}
One may now raise the question that entire analysis has been carried out by writing the state in the canonical form and the corresponding operators in the corresponding basis. So, if we are given a state in any other basis, will the analysis still hold? The fact, that any three qubit pure state can be written down in the canonical form is an existence proof that in principle one can always apply some local unitary operators to convert a state from any basis, in particular computational basis, to canonical-form basis and vice versa. 
We will now argue that these theorems will hold in any basis with suitably transformed operators.  We have to find the 
particular local unitary operation that connect two sets of basis vectors and write  those operator in that basis.
 Let us consider that we are given a three-qubit pure state : $\ket{\psi}=\sum\limits_{i,j,k=0}^1t_{ijk}\ket{ijk}$ in computational basis. Now, given this state, we can in principle always transform it to the canonical form \cite{acin2}. Only requirement is that one has to judiciously choose the local unitary operators. Following two examples will clarify this issue.
Suppose, we have been given a state in computational basis:
\begin{equation}\label{computational_basis}
\ket{\psi}_c=\frac{1}{2}(\ket{000}_c+\ket{011}_c+\ket{100}_c+\ket{111}_c),
\end{equation}
where $c$ represents that the state is in computational basis. Clearly the given state is not in canonical form. But it can be converted to one having the canonical form using local unitaries. For that we have to follow the prescription mentioned by Ac\'{i}n {\it et. al.} in \cite{acin2}. Doing the necessary calculations we have found that the unitary operators  that have to act on  the first,
second and third qubits are,\\
$
U1=\frac{1}{\sqrt{2}}\begin{pmatrix}
-1 & 1 \\
1 & 1
\end{pmatrix},
\;
U2=\begin{pmatrix}
1 & 0 \\
0 & 1
\end{pmatrix}
\;
\mbox{and}
\; U3=\begin{pmatrix}
1 & 0 \\
0 & 1
\end{pmatrix}.
$
Then by applying the operator $U=U1\otimes U2\otimes U3$ on $\ket{\psi}_c$, we get the state in the canonical-form basis as
\begin{equation}\label{acin_basis}
\ket{\psi}_a=\frac{1}{\sqrt{2}}(\ket{100}_a+\ket{111}_a),
\end{equation}
where, $a$ denotes that the state is in the canonical-form basis 
Now we can calculate the expectation values of those operators given in the previous sections. We will find that $\langle  O \rangle_{{\psi}_a}=0$, $\langle  O_1 \rangle_{{\psi}_a}=0$, $\langle  O_2 \rangle_{{\psi}_a}=0$ and $ \langle  O_3 \rangle_{{\psi}_a}=-2$. So the state is biseparable in $1$ and $23$ bipartition. Now to verify this result in computational basis we have to rotate these observables by inverse of $U$. $U^{-1}=U^\dagger=U1\otimes U2\otimes U3=U$. The transformed observables are $O_t=UOU^\dagger$,  $O_{1t}=UO_1U^\dagger$, $O_{2t}=UO_2U^\dagger$ and $O_{3t}=UO_3U^\dagger$. If we calculate the expectation values of thees operator on state $\ket{\psi}_c$, we find $\langle  O_t \rangle_{{\psi}_c}=0$, $\langle  O_{1t} \rangle_{{\psi}_c}=0$, $\langle  O_{2t} \rangle_{{\psi}_c}=0$ and  $\langle  O_{3t} \rangle_{{\psi}_c}=-2$. Hence, the state is $1|23$ biseparable. 

Let us consider another state in computational basis :
\begin{equation}\label{computational_basis1}
\ket{\phi}_c=\frac{1}{\sqrt{3}}(e^{i \theta}\ket{000}_c+\ket{011}_c-\ket{100}_c).
\end{equation}
To transform it in Ac\'{i}n's canonical form following local unitaries are required\\
$U1=\begin{pmatrix}
0 & -1 \\
1 & 0
\end{pmatrix},
\;
U2=\begin{pmatrix}
1 & 0 \\
0 & 1
\end{pmatrix}
\;
\mbox{and}
\; U3=\begin{pmatrix}
1 & 0 \\
0 & 1
\end{pmatrix}.$
Hence, $U=U1\otimes U2\otimes U3$. Applying this operator to
the state, we can get the final state as,
\begin{equation}\label{acin_basis1}
\ket{\phi}_a=\frac{1}{\sqrt{3}}(\ket{000}_a+e^{i \theta}\ket{100}_a+\ket{111}_a).
\end{equation}
For this state, we find that $\langle  O \rangle_{{\phi}_a}=\frac{4}{3}$, $\langle  O_1 \rangle_{{\phi}_a}=0$, $\langle  O_2 \rangle_{{\phi}_a}=0$ and $\langle  O_3 \rangle_{{\phi}_a}=-\frac{4\cos\theta}{3}$. So the state is in GHZ class. Now to get these results in computational basis we rotate these observables by $U^{-1}=U^\dagger=U1^\dagger\otimes U2 \otimes U3=U_I$. Similarly, we transform these observables by $U_I$, as shown in previous example. Now we find that $\langle  O_t \rangle_{{\phi}_c}=\frac{4}{3}$, $\langle  O_{1t} \rangle_{{\phi}_c}=0$, $\langle  O_{2t} \rangle_{{\phi}_c}=0$ and  $\langle  O_{3t} \rangle_{{\phi}_c}=-\frac{4\cos\theta}{3}$. Hence, the results are consistent. In the examples above,  it was important
to know the state to determine suitable unitary transformations. 


\section{Case of mixed states}
The case of mixed states is more involved. There is no closed from of tangle. But one can find a lower bound on the tangle for a three-qubit mixed state. For a mixed state $C_{A(BC)}^2=2(1-\mbox{Tr}\rho_A^2)$ is no longer valid. Here $\rho_A$ is the density matrix of a subsystem of the three-qubit state $\rho$. Instead we have to consider the convex roof optimization of all the pure states as follows
\begin{equation}\label{convex root}
C_{A(BC)}^2(\rho)=\inf_{p_i,\ket{\psi_i}}\sum_ip_iC_{A(BC)}^2(\ket{\psi_i}),
\end{equation} 
where $\rho=\sum_ip_i\ket{\psi_i}\bra{\psi_i}$. But it is a formidable task. Instead of finding this, one can find a lower bound on $C_{A(BC)}^2(\rho)$  easily.
It has been shown in   \cite{mintert} that this lower bound is given as -- $C_{A(BC)}^2(\rho)|_{LB}=2(\mbox{Tr}\rho^2-\mbox{Tr}\rho_A^2)$. By substituting this in the expression of tangle in Eq. (\ref{tangle}), one can find the lower bound on tangle. However, this $\tau^{LB}(\rho)$ is not always invariant under the permutation of $A$, $B$ and $C$. Hence, for the case of mixed states, it is reasonable to use the average over all the permutations of $A$, $B$ and $C$ and calculate  $\tau^{LB}(\rho)$ as follows \cite{mintert,farias}
\begin{equation}\label{tangle lower bound}
\bar{\tau}^{LB}=\frac{1}{6}\sum_{\{ABC\}}\Big(C_{A(BC)}^2|_{LB}-C_{AB}^2-C_{AC}^2\Big).
\end{equation}  
Let's take an example of a mixed state which is a mixture of a GHZ state and a W state
\begin{equation}\label{mixture ghz w}
\rho=p\ket{GHZ}\bra{GHZ}+(1-p)\ket{W}\bra{W}.
\end{equation}
For this state we compare graphically our observable measure of tangle for pure state, i.e. $\frac{\langle O\rangle^2}{4}$ with the lower bound of tangle as given in Eq. (\ref{tangle lower bound}).
\begin{figure}[h]
\centering
\includegraphics[scale=0.65]{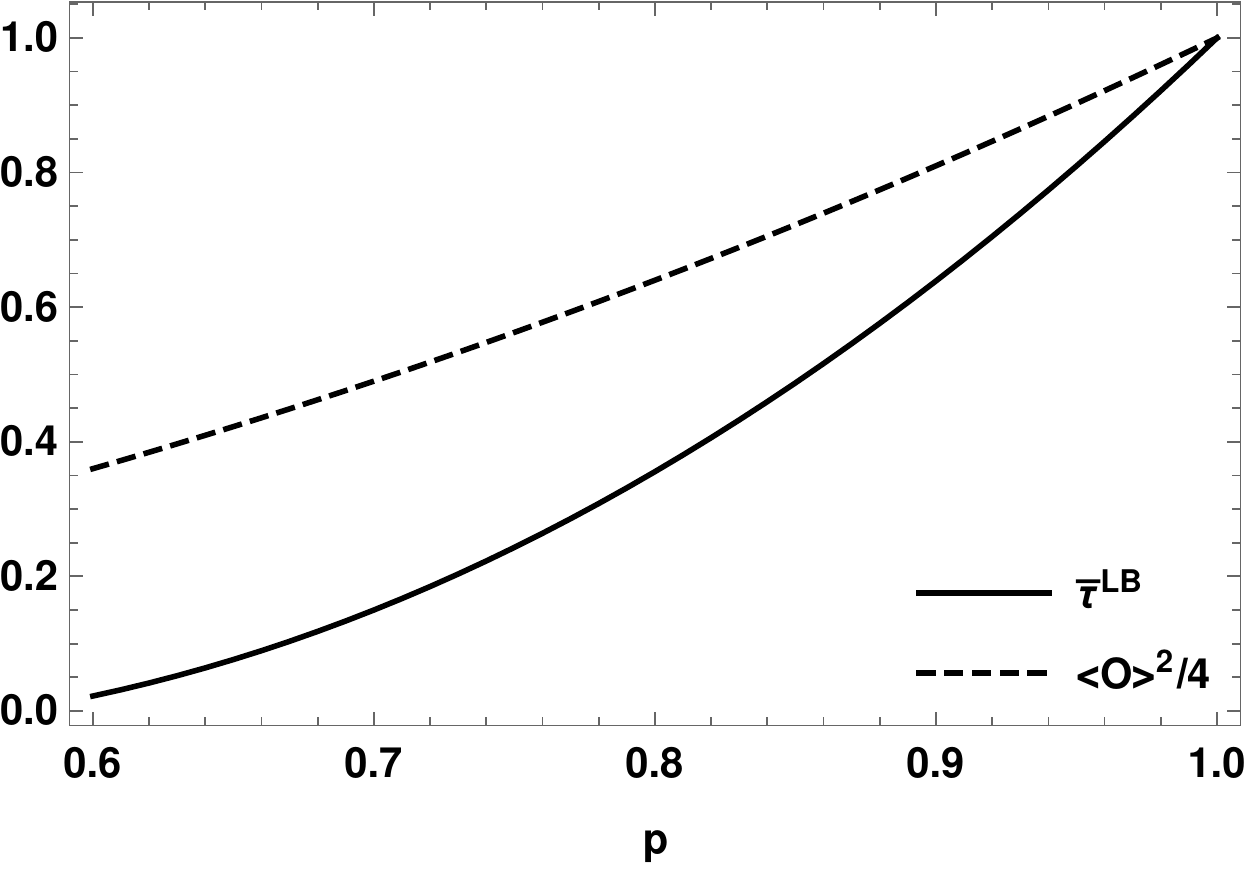}
\caption{Comparison between lower bound of tangle $(\bar{\tau}^{LB})$ and $\frac{\langle O\rangle^2}{4}$ with the variation of $p$ for the state given in Eq. (\ref{mixture ghz w}).}
\label{fig1}
\end{figure}
 As $p$ increases, the state becomes more pure and the values of $(\bar{\tau}^{LB})$ and $\frac{\langle O\rangle^2}{4}$ approach each other.
In \cite{farias}, Far\'{i}as et. al. considered a very interesting class of three-qubit mixed states which can be obtained as follows. First they prepare a two qubit Bell state $\ket{\phi^+}_{AB}=\frac{1}{\sqrt{2}}(\ket{00}+\ket{11})$. Then they let the second qubit interact with the environment. The interaction can be described by a phase damping channel
\begin{eqnarray}
&&\ket{0}_B\ket{0}_E\rightarrow\ket{0}_B\ket{0}_E\\
&&\ket{1}_B\ket{0}_E\rightarrow\sqrt{1-p}\ket{1}_B\ket{0}_E+\sqrt{p}\ket{0}_B\ket{1}_E,
\end{eqnarray} 
where $p$ is the channel parameter. This phase damping interaction prepares a tripartite state,
\begin{equation}\label{pd tripartite state}
\ket{\phi}_{ABE}=\frac{1}{\sqrt{2}}\Big(\ket{000}+\sqrt{1-p}\ket{110}+\sqrt{p}\ket{111}\Big),
\end{equation}
where initially the environment state is $\ket{0}$. In \cite{farias}, authors experimentally prepared this kind of state with some purity. We can represent it by adding some white noise with $\ket{\phi}_{ABE}$ as
\begin{equation}
\rho=m\ket{\phi}_{ABE}\bra{\phi}+\frac{1-m}{8}\mathbb{I},
\end{equation}
where $\mathbb{I}$ is the eight dimensional identity matrix. For purity equals to $0.92$ or $m\approx0.95$, we compare numerically our tangle measure for pure state, i.e., $\frac{\langle O\rangle^2}{4}$ with the lower bound of tangle. From the FIG. \ref{fig2}, we see that $\frac{\langle O\rangle^2}{4}$, i.e., our measure of tangle is just above the value of lower bound of tangle. Similar results can be obtained for any other class of mixed state as well.
\begin{figure}[h]
\centering
\includegraphics[scale=0.65]{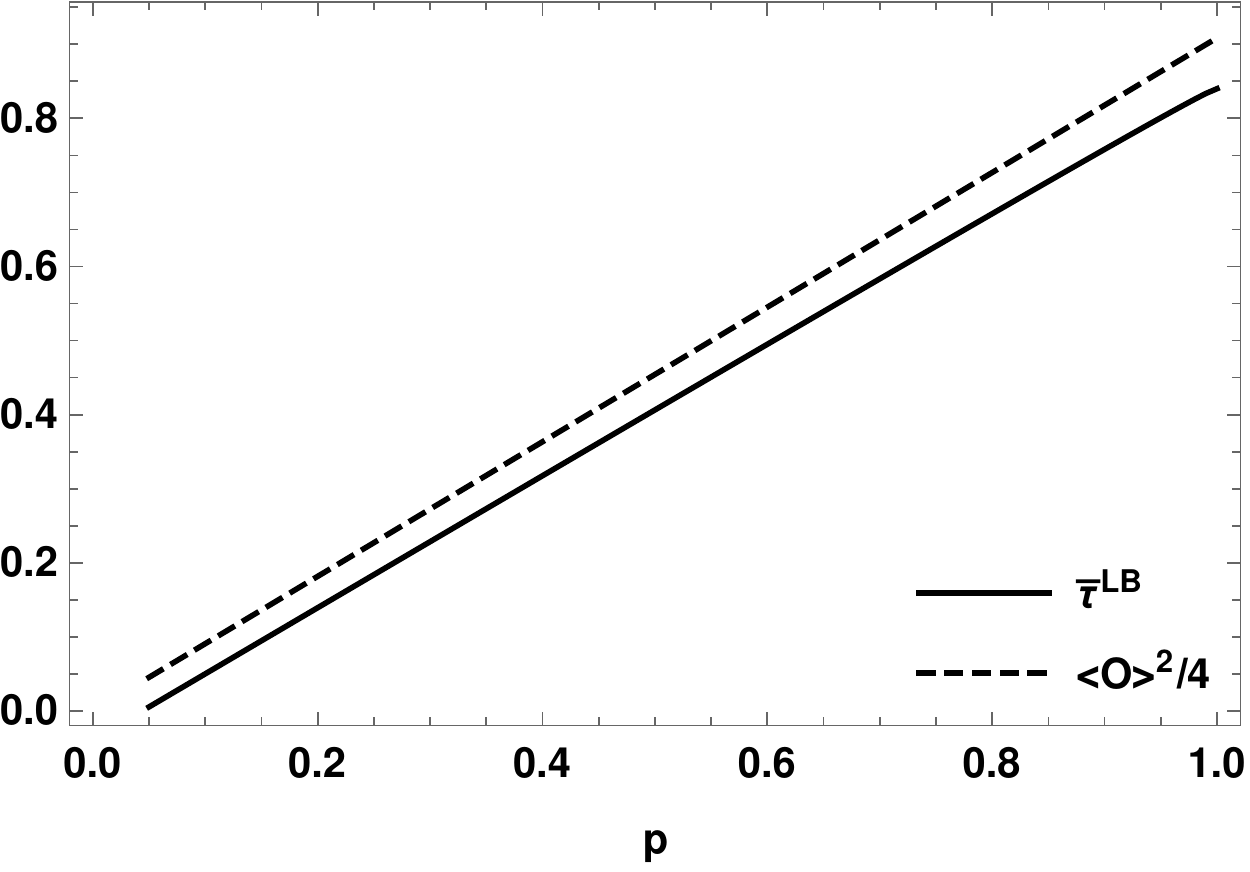}
\caption{Comparison between lower bound of tangle $(\bar{\tau}^{LB})$ and $\frac{\langle O\rangle^2}{4}$ with the variation of $p$.}
\label{fig2}
\end{figure}


\section{Experimental measure of fidelity for a teleportation scheme}
In this section, we will discuss a teleportation scheme using a
three-qubit pure state as studied earlier in \cite{lee}. The
teleportation scheme is as follows: Let us consider a three-qubit
pure entangled state shared by three parties $i,j$ and $k$. We
make an orthogonal measurement on the $k^{\rm th}$ qubit and consider the
joint state of the system $i$ and $j$. Using this joint state as a
resource state, one can teleport a single qubit state. The
faithfulness of this teleportation scheme depends on the single
qubit measurement on $k^{\rm th}$ qubit and the compound state of the
system $i$ and $j$. In \cite{lee}, authors introduced a new
quantity called partial tangle, which is defined as,
\begin{equation}\label{partial tangle}
\tau_{ij}=\sqrt{C_{i(jk)}^2-C_{ik}^2},~i\neq j\neq k~
{\rm and}~i,j,k=1,2,3.
\end{equation}
The partial tangles for the state given in (\ref{three-qubit state}) are
\begin{eqnarray}\label{partial tangles}
&&\tau_{12}=2\lambda_0\sqrt{\lambda_3^2+\lambda_4^2},\nonumber\\
&& \tau_{23}=2\sqrt{\lambda_0^2\lambda_4^2+\lambda_1^2\lambda_4^2+
\lambda_2^2\lambda_3^2-2\lambda_1\lambda_2\lambda_3\lambda_4
\cos\theta},\nonumber\\
&&\tau_{31}=2\lambda_0\sqrt{\lambda_2^2+\lambda_4^2}.
\end{eqnarray}
They showed that these partial tangles are related to singlet
fraction $f_k$ and maximum teleportation fidelity $F_k$. Here index $k$
just indicates that the measurement is done on the   $k^{\rm th}$ qubit.
The relation is
as follows,
\begin{equation}\label{fidelity and partial tangle}
\tau_{ij}=3F_k-2=2f_k-1.
\end{equation}

We will now provide the explicit relationship between the partial
tangles and the expectation value of the operators $O$, $O_{1}$,
$O_{2}$, $O_{4}$ and $O_{5}$. The operators $O_{4}$ and $O_{5}$
will be defined in this section. Since partial tangle is related
with singlet fraction and teleportation fidelity, we may
measure the teleportation fidelity experimentally for the
teleportation scheme given in \cite{lee}.

Let us define two new operators as
\begin{eqnarray}\label{O4 O5}
&& O_4=2(\sigma_z\otimes\sigma_y\otimes\sigma_y),\nonumber\\
&&O_5=2(\sigma_z\otimes\sigma_y\otimes\sigma_x).
\end{eqnarray}
Operator $O_4$ given in (\ref{O4 O5}) can be obtained from the Mermin operator (\ref{mermin operator}) by choosing the unit vectors as $\hat{a}_1=(0,0,1)$, $\hat{a}_2=(0,1,0)$, $\hat{a}_3=(0,1,0)$, $\hat{b}_1=(0,0,-1)$, $\hat{b}_2=(0,1,0)$ and $\hat{b}_3=(0,1,0)$. Similarly $O_5$ can be obtained by choosing the unit vectors as $\hat{a}_1=(0,0,1)$, $\hat{a}_2=(0,1,0)$, $\hat{a}_3=(1,0,0)$, $\hat{b}_1=(0,0,-1)$, $\hat{b}_2=(0,1,0)$ and $\hat{b}_3=(1,0,0)$.

The expectation value of the above observables for the state in (\ref{three-qubit state}) are
\begin{eqnarray}\label{exp O4 O5}
&&\langle O_4 \rangle_{\psi}
=-4(\lambda_2\lambda_3-\lambda_1\lambda_4\cos\theta),\nonumber\\
&&\langle O_5 \rangle_{\psi}=4\lambda_1\lambda_4\sin\theta.
\end{eqnarray}

After a few steps of calculation, we can show that,
\begin{eqnarray}\label{partial tangle observable}
&& \tau_{12}=\frac{1}{2}\sqrt{\langle O \rangle_{\psi}^2+
\langle O_1 \rangle_{\psi}^2}=3F_3-2,\nonumber\\
&& \tau_{23}=\frac{1}{2}\sqrt{\langle O \rangle_{\psi}^2+
\langle O_4 \rangle_{\psi}^2+\langle O_5 \rangle_{\psi}^2}=3F_1-2,\nonumber\\
&&\tau_{31}=\frac{1}{2}\sqrt{\langle O \rangle_{\psi}^2+ \langle
O_2 \rangle_{\psi}^2}=3F_2-2.\label{pet}
\end{eqnarray}
We note that the operators $O$, $O_{1}$, $O_{2}$, $O_{4}$ and
$O_{5}$ are observables and hence their expectation values are
measurable quantities. Since the teleportation fidelities are
related with some functions of these expectation values as shown
in (\ref{pet}), so we can say that the teleportation fidelities
for the teleportation scheme described in \cite{lee} may be
measured  experimentally.\\

From (\ref{exp O4 O5}) and  (\ref{partial tangle observable}),
we can draw following conclusions:

\begin{enumerate}
\item[1.]  If all the partial tangles are equal to zero then the state is a separable one. This is because
the expectation value of $O_4$ and $O_5$ are also zero for a separable state.

\item[2.]  If at least one partial tangle is equal to zero, then the three-qubit state is a biseparable state.

\item[3.]  If each partial tangle is not equal to zero then the state is a three-qubit genuine entangled state.\\

\end{enumerate}

\textit{Lemma 4:} Any pure three-qubit genuinely entangled state
is useful in the teleportation scheme of  \cite{lee}.\\

\textit{Proof :} Three-qubit genuinely entangled states consist of
GHZ-class and W-class. We will prove the proposition by taking
these two classes separately.
The relation in (\ref{fidelity and partial tangle}) can be written
as
\begin{equation}\label{fidelity and partial tangle 1}
F_k=\frac{2}{3}+\frac{\tau_{ij}}{3}.
\end{equation}
Case-I: For GHZ-class states, $\tau_{ij}>0$.This can be compared for If $\tau_{ij}>0$,
$F_k>\frac{2}{3}$ \cite{popescu}. Therefore the resource state
consisting of qubits $i$ and $j$ is suitable for teleportation. In
this case, the partial tangle is nonzero and so $\langle O
\rangle_{\psi}$ is also nonzero. Hence, from (\ref{partial tangle
observable}),
it is clear that $\tau_{ij}>0$. Therefore, $F_k$ is always greater than $\frac{2}{3}$. \\

Case-II: For W-class states, $\tau_{ij}=0$ and hence $\langle O
\rangle_{\psi}=0$. Therefore, for these class of states, the
equations (\ref{partial tangle observable}) reduces to
\begin{eqnarray}\label{partial tangle observable W}
&& \tau_{12}=\frac{1}{2}\langle O_1 \rangle_{\psi},\nonumber\\
&& \tau_{23}=\frac{1}{2}\sqrt{\langle O_4 \rangle_{\psi}^2+\langle O_5 \rangle_{\psi}^2},\nonumber\\
&&\tau_{31}=\frac{1}{2}\langle O_2 \rangle_{\psi}.
\end{eqnarray}
From (\ref{P}) it can be seen that $4 \tau_{12}\tau_{31}=P$. For
W-class states, $\tau_{12}\neq 0$ and $\tau_{31}\neq 0$ as $P\neq
0$. Hence, $F_3$ and $F_2$ are greater than $\frac{2}{3}$. Thus it
remains to see the remaining partial tangle $\tau_{23}$, which is
related with $\langle O_4 \rangle_{\psi}$ and $\langle O_5
\rangle_{\psi}$. From equation (\ref{exp PQ}), for W-class
states, $\lambda_0$, $\lambda_2$ and $\lambda_3$ can not be zero
simultaneously. Equation (\ref{O4 O5}) ensures that $\langle O_4
\rangle_{\psi}$ is nonzero. Therefore, $\tau_{23}\neq0$ and
$F_1$ is greater than $\frac{2}{3}$. Thus for the teleportation
scheme of \cite{lee}, all states in W-class are useful for
teleportation. This completes the proof.


\section{Conclusion}
We have constructed a set of operators, which can be used to
distinguish six SLOCC inequivalent classes of entanglement present
in pure three-qubit states. These operators contain only Pauli
matrices and hence are easily implementable in experiments. So, it may possible to
 detect the type of entanglement present in a three-qubit pure
state experimentally. Although we constructed theses operators
in the canonical-form basis, a suitably transformed set of operators
will work in any basis. In this sense, our results are independent
of the choice of the basis. However, the construction of suitable
transformations may require the knowledge about the state.
We have also considered a few mixed states and showed
graphically that our measure of tangle for pure states approaches
minimum of the tangle as the state becomes more pure. In the one class of mixed
states, that we considered, our measure of tangle is just above the lower bound on tangle.
This is because the purity of the states is quite high. In another case, our measure of tangle
approaches the lower bound, as the state becomes more pure. This shows that
our measure of tangle
works quite well for some classes of mixed state.
Also we have shown that the operators defined here can be used to
measure the fidelity of a teleportation scheme introduced in
\cite{lee}. We believe that, there are other such applications,
where we can use our operators effectively.

\vspace{0.5cm}

\noindent {\bf  Note:} It has come to our notice that an
 experiment has been performed \cite{experiment} based on the proposals in this paper.

\end{document}